# Bayesian Conditional Cointegration

**Chris Bracegirdle** C.BRACEGIRDLE@CS.UCL.AC.UK
**David Barber** D.BARBER@CS.UCL.AC.UK
Centre for Computational Statistics and Machine Learning, University College London, Gower Street, London

## Abstract

Cointegration is an important topic for time-series, and describes a relationship between two series in which a linear combination is stationary. Classically, the test for cointegration is based on a two stage process in which first the linear relation between the series is estimated by Ordinary Least Squares. Subsequently a unit root test is performed on the residuals. A well-known deficiency of this classical approach is that it can lead to erroneous conclusions about the presence of cointegration. As an alternative, we present a framework for estimating whether cointegration exists using Bayesian inference which is empirically superior to the classical approach. Finally, we apply our technique to model segmented cointegration in which cointegration may exist only for limited time. In contrast to previous approaches our model makes no restriction on the number of possible cointegration segments.

## 1. Introduction

Cointegration is an important concept in time-series analysis and relates to the fact that the differences between two time-series may be more predictable than any individual time-series itself. We make two contributions to this area—first, we formulate determining whether two series are cointegrated as an instance of Bayesian inference in an associated probabilistic model. This enables us to bring modern statistical and machine learning techniques to the analysis of cointegration and also provides a useful conceptual framework to describe cointegration. The resulting regression estimation algorithm is shown to be more robust than least-squares estimation to spurious results. This is we believe an important result that establishes a new and coherent approach to the long-standing problem of inconsistency in establishing cointegration. Second, in practice two series may only be intermittently cointegrated—that is, they are only cointegrated over shorter segments of time. To date the identification of these segments has been attempted with rather limiting assumptions such as the presence of either only two or three segments (Gregory & Hansen, 1996; Hatemi-J, 2008). To address this we phrase the problem as inference in a corresponding changepoint model, placing no limitation on the number of segments. We develop an efficient specialised inference scheme for this model and demonstrate its practical applicability on real-world problems.

### 1.1. Cointegration

Whilst individual time-series may not be predictable, the relationships between two time-series may be more predictable. For example, series $x_{1:T}$, $y_{1:T}$ formed by

$$x_{t+1} = x_t + \epsilon_{t+1}, \qquad y_{t+1} = y_t + \epsilon_{t+1}, \qquad \epsilon_t \sim \mathcal{N}(0, 1)$$

both follow an unpredictable random walk. However, the future difference $x_{t+1} - y_{t+1} = x_t - y_t$, is perfectly predictable given knowledge of the current difference. In the economics and finance literature such instances are common; for example underlying mechanisms such as limited money supply may forge dependencies between time-series (Dickey et al., 1991). It is of significant interest in finance to find pairs of asset price series that are cointegrated—such estimation underpins one of the classical statistical arbitrage strategies known as 'pairs trading'.

From an individual series $x_{1:T}$ we can form a new series $x^1_{2:T}$ by taking the difference $x^1_t = x_t - x_{t-1}$. Repeating this differencing process $d$ times, a series is order $d$ integrated, written $I(d)$, if the series $x^d_t$ formed from repeatedly taking the difference $d$ times yields a stationary series $I(0)$. In our example above, both $x$ and $y$ are each $I(1)$ and hence integrated. As we showed, there is a linear combination that is $I(0)$. More generally, two series $x_{1:T}$ and $y_{1:T}$ are cointegrated if they are each individually integrated and a linear combination of the two is integrated with a lower order.

For our purposes, we define $x_t$ and $y_t$ to be cointegrated within a specified time segment if there is a linear regression relationship between the two variables that forms a stationary process. For the most simple case of a single regressor, we write such a relationship as

$$\begin{aligned} y_t &= \alpha + \beta x_t + \epsilon_t \\ \epsilon_t &= \phi \epsilon_{t-1} + \eta_t, \quad \eta_t \sim \mathcal{N}\big(0, \sigma^2\big), \quad |\phi| < 1 \end{aligned}$$

where $\alpha$ represents a constant, and $\beta$ the regression coefficient. Here $\epsilon_{1:T}$ forms a mean-reverting, stationary process. For brevity, we focus attention on estimation in this model, however more flexible cases can easily be considered, including the case with a trend coefficient $\gamma$ in which the regression is written

$$y_t = \alpha + \beta x_t + \gamma t + \epsilon_t$$

with the same autoregression for $\epsilon_t$. Adapting our technique to more complex models with exogenous variables, and to the vector case, is also possible.

Testing for and estimating a cointegration relationship is classically a two-step process (Granger, 1986). Firstly, the regression equation is estimated based on a simple ordinary least squares (OLS) fit to minimise $\sum_t (y_t - \alpha - \beta x_t)^2$ (which we show is equivalent to maximum likelihood (ML) parameter optimisation assuming that $\phi = 0$ in a corresponding model). Subsequently, a test for a unit root in the residuals is performed using the Dickey-Fuller test, see for example Harris & Sollis (2003), which tests the hypothesis that $\phi = 1$ against the alternative $|\phi| < 1$ (for $|\phi| < 1$ the series is stationary and non-stationary otherwise). In the case that $\phi = 1$, it is well-known that OLS can deliver a spurious regression (Granger & Newbold, 1974), and this problem is not limited to cointegration. The classical approach is conceptually undesirable since it makes strong, potentially conflicting assumptions about the data: in the regression part, the residuals are effectively assumed uncorrelated whilst in the subsequent unit root test, they may be determined not to be (under the null hypothesis of the test, the regression was spurious). In this work we relax the assumptions placed on $\phi$ during parameter estimation to reduce the impact of any conflicting assumptions.

## 2. Modelling Cointegration

Our approach is to form a generative model of observations $p(y_{1:T}|x_{1:T}, \theta)$, where $\theta = \{\alpha, \beta, \sigma^2, \phi\}$. First, we form a generative model on observations[1] $y_{1:T}$ and latent variables $\epsilon_{1:T}$

$$p(y_{1:T}, \epsilon_{1:T}|x_{1:T}) = \prod_t p(y_t|x_t, \epsilon_t)\, p(\epsilon_t|\epsilon_{t-1}), \quad \epsilon_0 = \emptyset$$

where[2]

$$p(y_t|x_t, \epsilon_t) = \delta(y_t - \alpha - \beta x_t - \epsilon_t)$$

and the transition for $\epsilon_t$ is given as

$$p(\epsilon_t|\epsilon_{t-1}) = \mathcal{N}\big(\epsilon_t \big| \phi \epsilon_{t-1}, \sigma^2\big).$$

The belief network (see Barber (2012) for an introduction) for this model is given in figure 1. The likelihood on the observations $y_{1:T}$ is given by

$$p(y_{1:T}|x_{1:T}) = \int_{\epsilon_{1:T}} p(y_{1:T}, \epsilon_{1:T}|x_{1:T}).$$

The integration distributes,

$$\begin{aligned} p(y_{1:T}|x_{1:T}) = \int_{\epsilon_{1:T-1}} \prod_{t=1}^{T-1} p(y_t|x_t, \epsilon_t)\, p(\epsilon_t|\epsilon_{t-1}) \\ \times \int_{\epsilon_T} p(y_T|x_T, \epsilon_T)\, p(\epsilon_T|\epsilon_{T-1}) \end{aligned}$$

and since we have a delta function, the final term evaluates to $\mathcal{N}\big(y_T - \alpha - \beta x_T | \phi \epsilon_{T-1}, \sigma^2\big)$. Iterating this yields a product of Gaussian terms

$$\prod_{t=2}^{T} \mathcal{N}\big(y_t - \alpha - \beta x_t \big| \phi\,(y_{t-1} - \alpha - \beta x_{t-1}), \sigma^2\big).$$

Finally, applying labels according to $\epsilon_t = y_t - \alpha - \beta x_t$,

$$p(y_{1:T}|x_{1:T}) = p(\epsilon_1) \prod_{t=2}^{T} \mathcal{N}\big(\epsilon_t \big| \phi \epsilon_{t-1}, \sigma^2\big) = p(\epsilon_{1:T}).$$

Hence the likelihood $p(y_{1:T}|x_{1:T})$ is equivalent to the likelihood on the Markov chain with 'observations' $\epsilon_t$ shown in figure 2. The log likelihood $\log p(\epsilon_{1:T})$ is

$$\log p(\epsilon_1) - \frac{1}{2\sigma^2} \sum_{t=2}^{T} (\epsilon_t - \phi \epsilon_{t-1})^2 - \frac{T-1}{2} \log 2\pi\sigma^2$$

[1]Note that we make no assumption about the underlying process $x$; we work with the conditional relationship $p(y_{1:T}|x_{1:T})$ not the joint $p(x_{1:T}, y_{1:T})$. As far as we are aware previous Bayesian approaches to cointegration make stronger assumptions about the underlying process, see for example Koop et al. (2006).

[2]The Dirac delta distribution $\delta(x-a)$ represents a degenerate probability density function with the property that $\int_x \delta(x-a)\, f(x) = f(a)$.

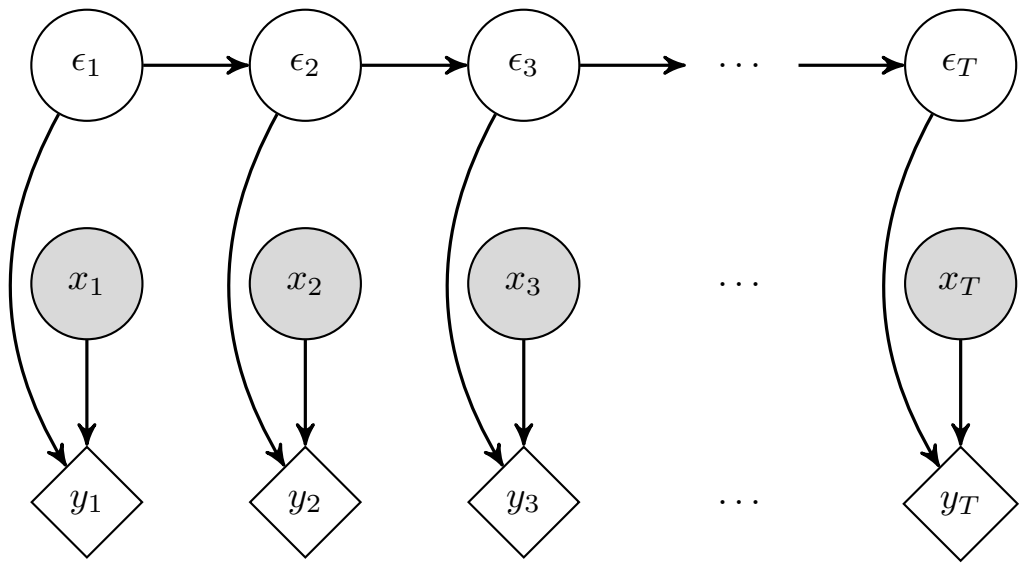

*Figure 1.* Belief network for the natural model for cointegration. Diamonds represent delta functions, shaded variables are in the conditioning set.

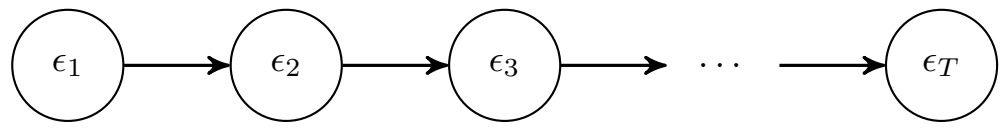

*Figure 2.* Belief network for the Markov chain model for $\epsilon_t$.

and when $\phi = 0$, maximising this degenerates to minimising the sum of squared residual terms

$$\sum_{t=2}^{T} \epsilon_t^2 = \sum_{t=2}^{T} \left(y_t - \alpha - \beta x_t\right)^2 .$$

OLS estimation for the regression parameters $\alpha$ and $\beta$ therefore corresponds to the ML solution of this model, on the assumption that $\phi = 0$. In the case that $\phi$ is non-zero, a different solution may be optimal in the ML sense. This is a source of potential inconsistency in the classical approach to cointegration testing.

## 3. Bayesian Cointegration

We can construct a model for estimating both $\phi$ and the parameters $\alpha, \beta, \sigma^2$ by considering $\phi$ to be a latent variable and providing a prior distribution. In this case, for cointegration we require $|\phi| < 1$, and we can encode this with a uniform distribution $p(\phi) = \mathcal{U}(\phi|\,(-1,1)) = \frac{1}{2}\,[\phi \in (-1,1)]$. We complete the model specification with a distribution for $\epsilon_1$ and form the joint model

$$p(y_{1:T}, \epsilon_{1:T}, \phi | x_{1:T}) = p(y_{1:T}|\epsilon_{1:T}, x_{1:T})\, p(\epsilon_{1:T}|\phi)\, p(\phi)$$

from which the marginal model

$$p(y_{1:T}, \phi | x_{1:T}) = p(\phi) \int_{\epsilon_{1:T}} p(y_{1:T}|\epsilon_{1:T}, x_{1:T})\, p(\epsilon_{1:T}|\phi)$$

is formed by integration. The posterior $p(\phi|x_{1:T}, y_{1:T})$ is then equivalent to $p(\phi|\epsilon_{1:T}) \propto p(\phi)\, p(\epsilon_{1:T}|\phi)$.

### 3.1. Inference of $\phi$

The process $\epsilon_{1:T}$ is stationary when each[3] $\langle \epsilon_t \rangle = 0$ and $\left\langle \epsilon_t^2 \right\rangle$ is constant, independent of $t$. According to the recurrence relation for $\epsilon_t$, the variance satisfies $\left\langle \epsilon_{t+1}^2 \right\rangle = \phi^2 \left\langle \epsilon_t^2 \right\rangle + \sigma^2$ and we can therefore ensure stationary of $\epsilon_{1:T}$ if $\left\langle \epsilon_1^2 \right\rangle = \sigma^2 / \left(1 - \phi^2\right)$. We then choose $p(\epsilon_1|\phi) = \mathcal{N}\left(\epsilon_1 \middle| 0, \sigma^2/\left(1-\phi^2\right)\right)$. The posterior $p(\phi|\epsilon_{1:T})$ is proportional to

$$p(\phi)\,\sqrt{1-\phi^2}\,\exp\frac{-1}{2\sigma^2}\left[\epsilon_1^2\left(1-\phi^2\right) + \sum_{t=2}^{T}\left(\epsilon_t - \phi\epsilon_{t-1}\right)^2\right]$$

and by completing the square we see that the posterior is given by a truncated Gaussian distribution with a prefactor (subject to normalisation),

$$p(\phi)\,\sqrt{1-\phi^2}\,\mathcal{N}\left(\phi \middle| \frac{\hat{e}_{12}}{\hat{e}_1}, \frac{\sigma^2}{\hat{e}_1}\right)$$

where

$$\hat{e}_{12} \equiv \sum_{t=2}^{T} \epsilon_t \epsilon_{t-1}, \qquad \hat{e}_1 \equiv \sum_{t=3}^{T} \epsilon_{t-1}^2 .$$

For the data likelihood, we write for $p(\epsilon_{1:T})$,

$$\left(2\pi\sigma^2\right)^{\frac{1-T}{2}} \exp -\frac{1}{2\sigma^2}\left(\sum_{t=1}^{T}\epsilon_t^2 - \frac{\left(\hat{e}_{12}\right)^2}{\hat{e}_1}\right) \times \frac{1}{\sqrt{\hat{e}_1}} \int_{-1}^{1}{}_{\phi}\, \tfrac{1}{2}\sqrt{1-\phi^2}\,\mathcal{N}\left(\phi \middle| \frac{\hat{e}_{12}}{\hat{e}_1}, \frac{\sigma^2}{\hat{e}_{12}}\right).$$

### 3.2. Estimating $\alpha, \beta, \sigma^2$

Our interest is to set the parameters $\theta = \left\{\alpha, \beta, \sigma^2\right\}$ based on maximising the likelihood

$$p(y_{1:T}|x_{1:T}, \theta) = \int_{\phi} p(\phi)\, p(\epsilon_{1:T}|\phi) .$$

Since $\phi$ is a latent variable, it is convenient to approach this using the Expectation-Maximisation algorithm. Writing $v$ for the observations and $h$ for the latent variables EM is based on the Kullback-Leibler bound

$$0 \le KL(p\|q) \equiv \left\langle \log\frac{q(h|v)}{p(h|v)} \right\rangle_{q(h|v)}$$

$$\log p(v) \ge \underbrace{\langle \log p(h,v) \rangle_{q(h|v)}}_{\text{energy}} - \underbrace{\langle \log q(h|v) \rangle_{q(h|v)}}_{\text{entropy}}$$

where $p$ is a variational distribution and $q$ a fixed posterior. EM corresponds to iteratively maximising this lower bound with respect to the variational distribution $p$, and using this as the next $q$ following inference. For this model, we replace $h \to \phi$, $v \to \epsilon_{1:T}$.

[3] Angled brackets represent expectation, $\langle x \rangle = \mathbb{E}\,[x]$.

The energy term[4] is the log of the model joint, and since we are interested in maximising with respect to the regression parameters $\alpha$ and $\beta$ the only relevant terms are given as

$$\langle \log p(\epsilon_1|\phi)\rangle_{q(\phi|\epsilon_{1:T})} + \sum_{t=2}^{T} \langle \log p(\epsilon_t|\epsilon_{t-1},\phi)\rangle_{q(\phi|\epsilon_{1:T})}$$

and up to a constant, this equals

$$\frac{-1}{2\sigma^2}\left[\epsilon_1^2\left\langle 1-\phi^2\right\rangle + \sum_{t=2}^{T}\left\langle (\epsilon_t - \phi\epsilon_{t-1})^2\right\rangle\right] - \frac{T}{2}\log 2\pi\sigma^2$$

where $q(\phi|\epsilon_{1:T}) = p\big(\phi|\epsilon_{1:T}, \theta^{\text{old}}\big)$ is the posterior from the previous iteration. The energy can be optimised by finding the stationary point by differentiating by $\alpha$ and $\beta$. Since

$$\left\langle (\epsilon_t - \phi\epsilon_{t-1})^2\right\rangle = \epsilon_t^2 - 2\epsilon_t\epsilon_{t-1}\left\langle\phi\right\rangle + \epsilon_{t-1}^2\left\langle\phi^2\right\rangle$$

the result is a system of linear equations for $\alpha$ and $\beta$ involving the first and second (non-central) moments of $\phi$ from the posterior $q(\phi|\epsilon_{1:T})$.

By differentiating the above energy term with respect to $\sigma^2$, we find that optimally

$$\widehat{\sigma}^2 = \frac{1}{T}\left[\epsilon_1^2\left\langle 1-\phi^2\right\rangle + \sum_{t=2}^{T}\left\langle (\epsilon_t - \phi\epsilon_{t-1})^2\right\rangle\right]$$

so the variance can be estimated once the new regression estimates $\alpha$ and $\beta$ have been found.

# 4. The Random Walk model

A primary concern in the economics literature is to test the hypothesis of cointegration for series. The analogue with our Bayesian generative model is to compare the likelihoods of the above cointegration model and a random walk (RW) model. The likelihood for $\epsilon_{1:T}$ a RW is calculated as

$$p(\epsilon_{1:T}|\phi = 1) = p(\epsilon_1)\prod_{t=2}^{T}\mathcal{N}\big(\epsilon_t|\epsilon_{t-1},\sigma^2\big)$$

For $p(\epsilon_1)$ we choose a wide-interval uniform distribution. The cointegration model and random walk model can then be compared according to the Bayes factor,

$$\frac{p(y_{1:T}|x_{1:T};\phi = 1)}{p(y_{1:T}|x_{1:T};|\phi| < 1)} = \frac{p(\epsilon_{1:T}|\phi = 1)}{p(\epsilon_{1:T}|\,|\phi| < 1)} \equiv \frac{l_{\text{RW}}}{l_{\text{C}}}$$

[4]Also called the expected completed data log likelihood.

**Algorithm 1** Bayesian cointegration testing

```
{α, β, σ²} ← LinearRegression(x_{1:T}, y_{1:T})
repeat
    ε_{1:T} ← y_{1:T} − α − β x_{1:T}
    {l_C, ⟨φ⟩, ⟨φ²⟩} ← CointInference(ε_{1:T}, σ²)
    {α, β, σ²} ← EM(x_{1:T}, y_{1:T}, ⟨φ⟩, ⟨φ²⟩)
until convergence
l_RW ← ∏_{t=2}^{T} N(ε_t | ε_{t−1}, σ̂²)
return cointegrated ← l_RW / l_C < threshold
```

where the numerator represents the 'null hypothesis' of a unit root in the residuals process. The denominator is simply the marginal likelihood for the Bayesian cointegration model given in section 3.1.

Whilst a fully-Bayesian approach to cointegration testing may be considered by placing prior distributions and seeking to integrate over the parameters $\alpha$, $\beta$, $\sigma^2$ separately for each model, here we restrict our analysis to point estimation for reasons of both simplicity and computational speed. Furthermore we wish to relate closely to the classical point estimate approach to cointegration estimation and testing. The purpose of this study is demonstrate the potential advantage of using a 'partial' Bayesian approach for $\phi$ alone.

## 4.1. Point Estimation

Both the numerator and denominator of the Bayes factor are functions in the variance $\sigma^2$, and we set $\sigma$ for each of the two models by ML. The estimate for $\sigma^2$ for the RW model is given according to

$$\widehat{\sigma}^2 = \frac{1}{T-1}\sum_{t=2}^{T}(\epsilon_t - \epsilon_{t-1})^2.$$

Note that here $\epsilon_t$ depends on $\alpha$ and $\beta$, and we now discuss possible ways to determine point estimates for these parameters. There are three primary options: (i) find the parameters by OLS and use these values for both the cointegration and RW models; (ii) find the parameters by ML in the cointegration model and use the same parameters for the RW model; and (iii) find the parameters for each of the cointegration and RW models by ML. Conceptually, our interest is to first *estimate* a cointegration relationship, and second to *test* whether this relationship is more likely to be a random walk—and for this reason, we consider the third option undesirable. Whilst the first option is computationally the simplest it suffers from the implicit assumption that $\phi = 0$. For this reason the second is our preferred method.

Algorithm 1 shows the calculation steps for the Bayesian cointegration test. The results in figure 3

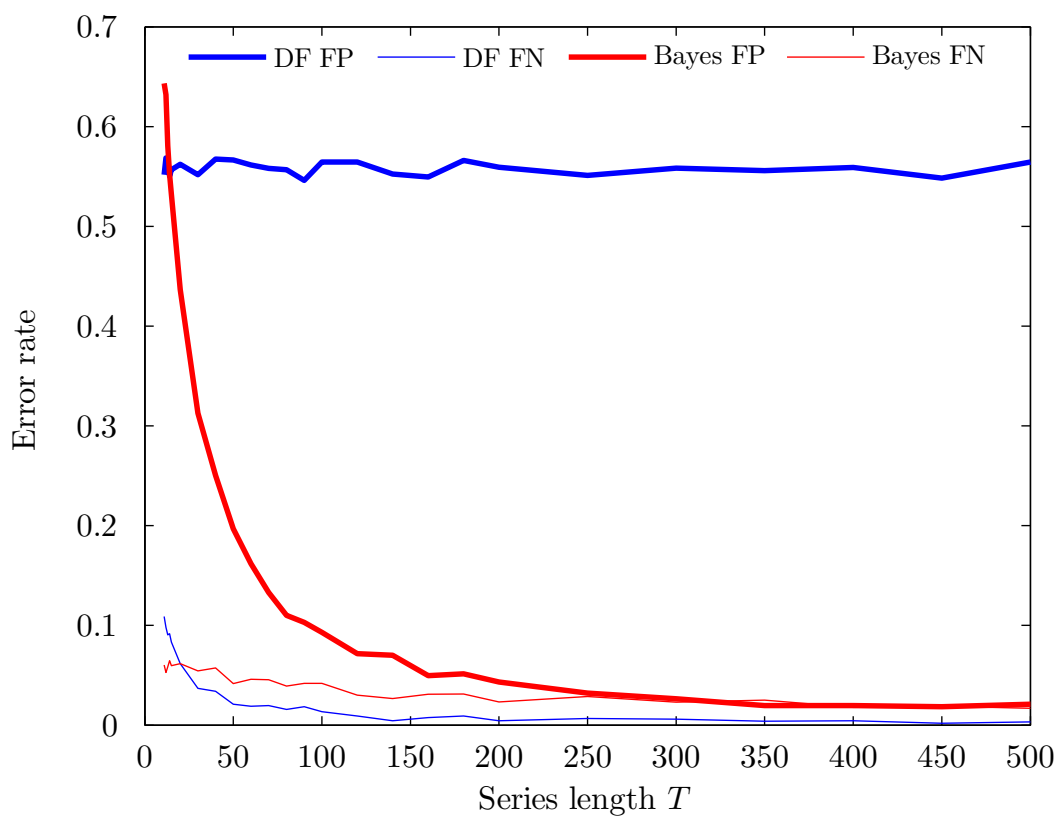

*Figure 3.* Plot of false positive (FP) and false negative (FN) rates (calculated as number of FP/FN outcomes divided by number of true negatives/positives respectively) for the classical OLS/DF approach (5% significance) and Bayesian EM/likelihood method (threshold $C = \exp 2$) for generated series. 5000 series of each length were simulated according to the generative model, uniformly split between $|\phi| < 1$ in the cointegrated case and $\phi = 1$ in the random walk case. The high rate of false positives for the classical test is striking: here cointegration was detected when the data were generated with $\phi = 1$. We attribute this to the fact that OLS is known to produce invalid estimates when the process is non-stationary: so called 'spurious regression'.

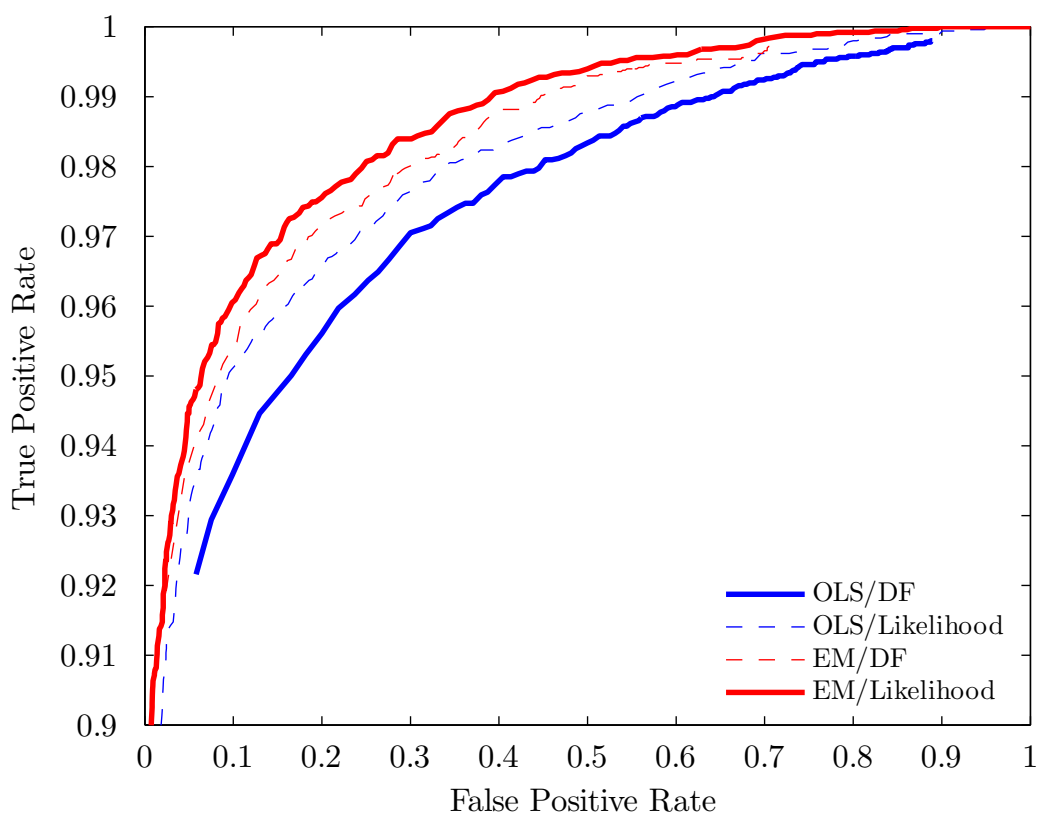

*Figure 4.* Plot of the receiver operating characteristic (ROC) curve for the cointegration estimation/testing combinations, for $T = 100$. 10,000 series were generated; we plot the true positive rate against the false positive rate for different values of the decision threshold. EM refers to estimation in the Bayesian cointegration model by ML; DF refers to a Dickey-Fuller test; Likelihood refers to a comparison of the likelihood in the cointegration model with the random walk model $\phi = 1$. Optimal classification occurs with true positive rate 100%, false positive rate 0%—the results show that OLS/DF is dominated by EM/Likelihood.

show that, compared with OLS estimation and unit root testing, our Bayesian technique is less likely to result in a spurious relationship for series of length $T > 20$. In figure 4 we show comparisons for differing decision boundaries[5].

## 5. Intermittent Cointegration

For some series, cointegration may only apply for certain segments. A good example is the Interconnector gas pipeline between Bacton, UK and Zeebrugge, Belgium, which allows gas to flow between the two countries, providing a direct link in the gas price at each end. When the pipeline closes each year for maintenance, the link between the prices is temporarily broken. It was shown by Kim (2003) that classical tests for cointegration may fail to detect a relationship even though, for a part of the series, there is such a relationship. The following model seeks to detect a cointegration relationship between two series, while allowing for regions when the relationship is in fact a random walk.

[5]Differing decision thresholds correspond to different widths of uniform prior for $\epsilon_1$ in the RW model since

$$\frac{l_{\mathrm{RW}}}{l_{\mathrm{C}}} < C \Leftrightarrow \frac{\tilde{l}_{\mathrm{RW}}}{l_{\mathrm{C}}} < 1, \quad \tilde{l}_{\mathrm{RW}} \equiv \frac{l_{\mathrm{RW}}}{C}.$$

Previous works in this area typically limit the number of regimes in which cointegration can occur to either only two or three segments (Gregory & Hansen, 1996; Hatemi-J, 2008). In contrast, for our model we allow time-varying $\phi_t$ arranged to be piecewise-constant with no *a priori* restriction on the number of piecewise-constant regions. The model will switch between the cointegrated case $|\phi_t| < 1$ and the alternative $\phi_t = 1$, corresponding to a random walk.

We use the binary state switch $i_t$ to denote whether there is a cointegration relationship at time $t$: $i_t = 1$ denotes regions of $\epsilon_t$ corresponding to a random walk; alternatively when $i_t = 0$, $\epsilon_t$ follows a stationary cointegration relation. In the random walk regions, we have $\phi_t = 1$, which is encoded in the prior distribution $p^1(\phi_t) = \delta(\phi_t - 1)$. We therefore specify

$$p(\phi_2|i_2 = 0) = \begin{cases} p^1(\phi_2) = \delta(\phi_2 - 1) & i_2 = 1 \\ p^0(\phi_2) = \mathcal{U}(\phi_2|\,(-1,1)) & i_2 = 0 \end{cases}$$

and for the piecewise-constant transition,

$$p(\phi_t|\phi_{t-1}, i_t, i_{t-1}) = \begin{cases} p^1(\phi_t) & i_t = 1 \\ \delta(\phi_t - \phi_{t-1}) & i_t = 0, i_{t-1} = 0 \\ p^0(\phi_t) & i_t = 0, i_{t-1} = 1. \end{cases}$$

Whilst the state transition $p(i_t|i_{t-1})$ and $p(i_2)$ can also in principle be learned on the basis of ML, we leave these quantities to be user specified. Note that whilst $\phi_t$ can change with time, for computational tractability we assume that the remaining parameters $\alpha$, $\beta$, $\sigma^2$ are fixed throughout time. The belief network representation is given in figure 5.

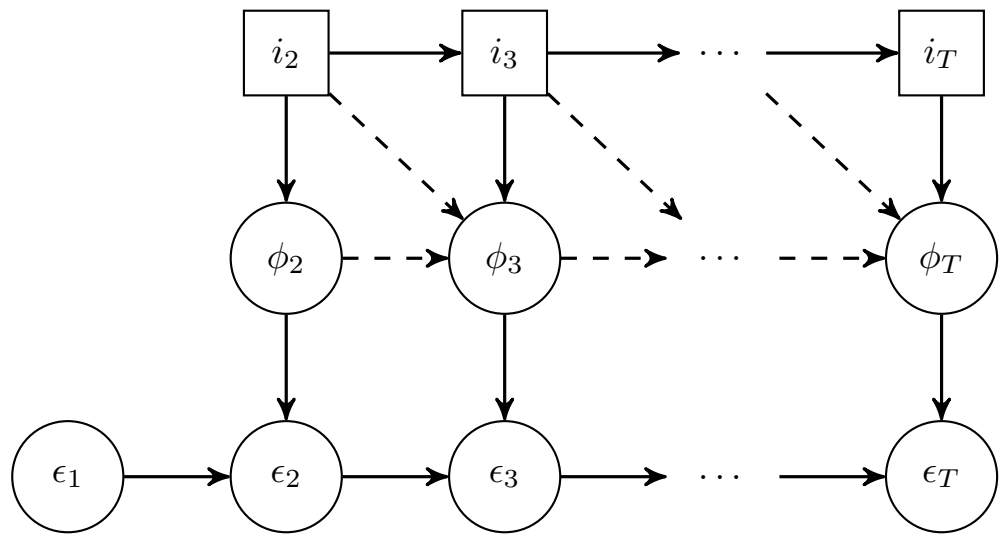

*Figure 5.* Belief network for the intermittent cointegration model. Dashed edges to $\phi_t$ are absent when $i_t = 1$.

The switching inference routine described in the following section relies on the recursive updates given in appendix A, in which we use the uniform prior for $\epsilon_t$ at the start of each cointegrated segment.

### 5.1. Inference of $i_{2:T}$, $\phi_{2:T}$

The regime-switching model with piecewise-constant $\phi_t$ is an example of a changepoint model, in which a change of regime drops the dependency of the continuous latent variable $\phi_t$ on the past. The model is therefore an example of the 'reset model' as set out by Bracegirdle & Barber (2011), and we apply such results here. Exact inference is then possible by applying recursions for filtering (calculating each $p(\phi_t, i_t|\epsilon_{1:t})$) and correction-smoothing (for $p(\phi_t, i_t|\epsilon_{1:T})$). For brevity, we show here only some of the key considerations; a full derivation is given in the supplementary material.

It is easy to show that both the filtered and smoothed posteriors for $\phi_t$ in the case that $i_t = 1$ degenerate in an intuitive way to

$$p(\phi_t|i_t = 1, \epsilon_{1:t}) = p(\phi_t|i_t = 1, \epsilon_{1:T}) = p^1(\phi_t)\,.$$

For the case $i_t = 0$, the posterior for $\phi_t$ is more complex because of the temporal dependency. For filtering, the posterior depends on the previous regime: in the event that $i_{t-1} = 1$, a new cointegration regime has begun, and a Gaussian term is contributed to the posterior in accordance with (2); otherwise, each component is updated as for the simple cointegration case (1). We therefore see that the posterior $p(\phi_t|i_t = 0, \epsilon_{1:t})$ is given as a truncated mixture of Gaussian components.

For correction smoothing, it can be difficult to derive a recursion analytically because the backwards-recursive step requires a 'dynamics reversal' term that is calculated with the filtered posterior in the denominator. In the case that the filtered posterior is a mixture distribution, the algebra is intractable. However, we appeal to the result of Bracegirdle & Barber (2011) that by indexing the components in the filtered posterior according to the number of observations since a switch to cointegration, the smoothing recursion can be written exactly. The resulting recursion shows that, as we may expect, a subset of the Gaussian components from $p(\phi_{t+1}|i_{t+1} = 0, \epsilon_{1:T})$ are contributed to the posterior $p(\phi_t|i_t = 0, \epsilon_{1:T})$ without change, along with the components from filtering $p(\phi_t|i_t = 0, \epsilon_{1:t})$ in the case that the cointegration regime ends at $t$.

The discrete filtering and smoothing components, $p(i_t|\epsilon_{1:t})$ and $p(i_t|\epsilon_{1:T})$, are also calculated as part of the recursions. The result is an algorithm for exact inference in this switching model that scales as $T^2$.

### 5.2. Likelihood

It is useful to evaluate the likelihood in order to (i) ensure that the likelihood is maximised, and (ii) provide a convergence criterion. Fortunately the likelihood is easy to calculate since the value is given by the product of normalisation constants from each step of the filtering recursion,

$$p(\epsilon_{1:T}) = p(\epsilon_1)\prod_{t=2}^{T} p(\epsilon_t|\epsilon_{1:t-1}) = p(\epsilon_1)\prod_{t=2}^{T} Z_t$$

where each $Z_t$ is calculated when filtering, see the supplementary material.

### 5.3. Learning

EM is again used for parameter estimation in this model. For this regime-switching problem, the latent variables are $h \to \{i_{2:T}, \phi_{2:T}\}$, and the observations remain as before, $v \to \epsilon_{1:T}$.

Terms relevant to the optimal solution are the sum of quadratic forms derived in section 3.2 with varying $\phi_t$,

$$\left\langle (\epsilon_t - \phi_t\epsilon_{t-1})^2 \right\rangle = \epsilon_t^2 - 2\epsilon_t\epsilon_{t-1}\left\langle \phi_t \right\rangle + \epsilon_{t-1}^2\left\langle \phi_t^2 \right\rangle.$$

By retaining the first and second (non-central) moments of each component found while filtering, the linear system of equations for the regression parameters can be solved exactly in this switching model, and the variance estimate can be updated accordingly.

### 5.4. Point Estimate of $\phi_t$

Whilst our model gives a distribution over $\phi_t$ for each timepoint, in order to be able to compare our approach

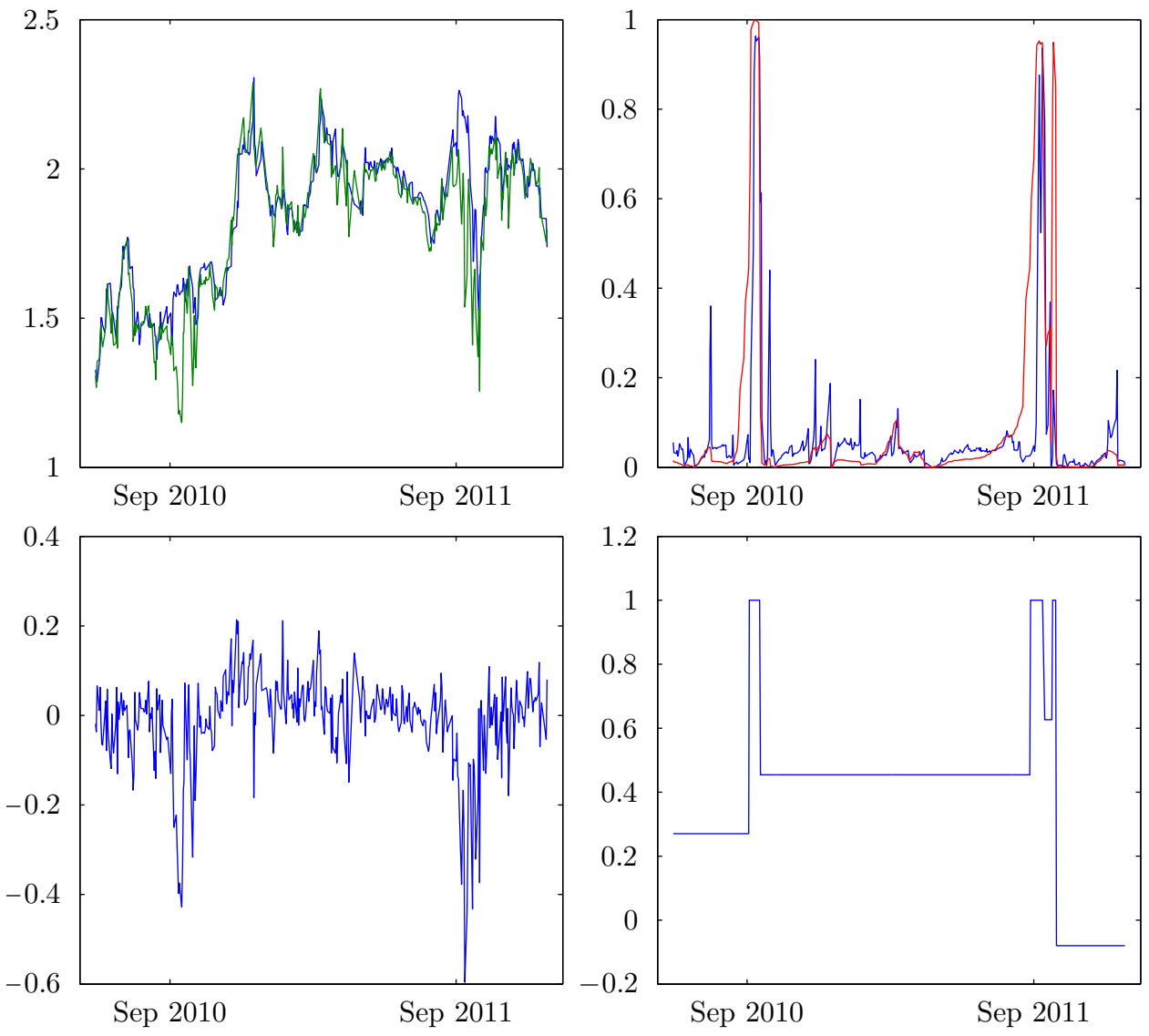

*Figure 6.* Results of learning for the Interconnector gas prices—maintenance occurs each September. We show (NW) the price series $x_t$, $y_t$; (SW) residuals $\epsilon_t$ calculated with the final estimates for $\alpha$, $\beta$; (NE) filtered (blue) and smoothed (red) posterior for random walk $i_t = 1$; and (SE) the maximum posterior marginal result for $\phi_t$.

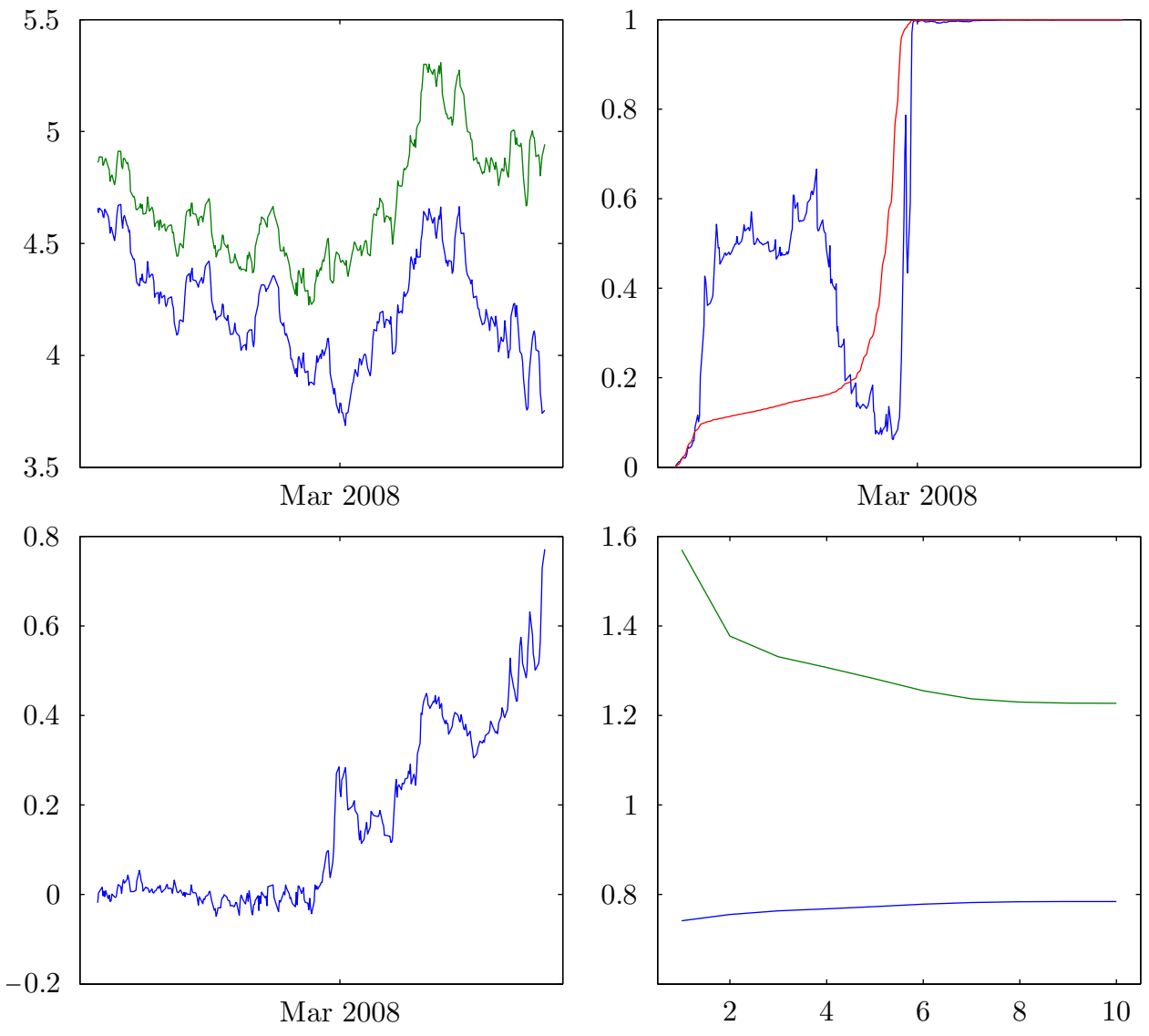

*Figure 7.* Results of learning for the Euro-area bond yields. We show (NW) the yield series $x_t$, $y_t$; (SW) residuals $\epsilon_t$ calculated with the final estimates for $\alpha$, $\beta$; (NE) filtered (blue) and smoothed (red) posterior for random walk $i_t = 1$; and (SE) the convergence of parameters $\alpha$, $\beta$—note the relative change in the estimate for intercept $\alpha$ (green).

to other intermittent cointegration methods that only give point estimates, we may take each

$$\widehat{i}_t = \arg\max_{i_t} p(i_t | \epsilon_{1:T})$$

and then calculate the posteriors $p\left(\phi_t \middle| \widehat{i}_{2:T}, \epsilon_{1:T}\right)$ to give the point estimates

$$\widehat{\phi}_t = \arg\max_{\phi_t} p\left(\phi_t \middle| \widehat{i}_{2:T}, \epsilon_{1:T}\right).$$

## 6. Experiments

We consider real-world applications for the intermittent cointegration model. For these experiments, we ran the inference and learning algorithm for the parameters $\theta = \{\alpha, \beta, \sigma\}$ initialised by OLS, and use the state transition distributions fixed at values designed to match our *a priori* belief about the cointegration regimes.

### 6.1. Gas Prices

The Interconnector, as noted in section 5, is a sub-sea natural gas pipeline: we took for $y$ UK gas prices[6] and for $x$ the Zeebrugge prices[7]. There are approximately 245 pricing days per year, and the pipeline closes annually for two weeks, so we specified the state transition distribution accordingly. The algorithm reaches convergence within four minutes with parameters $\alpha$, $\beta$ that show cointegration between the annual maintenance period, see figure 6.

### 6.2. Euro-area Bond Yields

We consider possible cointegration between Greek and German 10-year benchmark bond yields[8] prior to the Euro-area financial crisis. We use the cointegration switching model to find regions of cointegration, which we expect to hold until the yield on Greek debt spiralled as Greece's debt burden took toll. The results are shown in figure 7; inference and learning takes roughly five minutes. This is an example for which, over the time window shown in the figure, the data do not show cointegration according to the classical test (Dickey-Fuller test shows p-value 0.927241, and does not reject the null hypothesis of random walk), but the intermittent cointegration model does show a segment of cointegration—the classical test passes in the detected region, and fails for the remainder.

Whilst our approach is both flexible and simple, there are limitations. In particular, the switching model requires that the regions of cointegration are governed by a time-invariant linear relationship. For example, in the case that deviations from the relationship in the random-walk segments are permanent, the model presented here would struggle to properly estimate a relationship since piecewise-constant values of the constant term $\alpha$ would be required. Such considerations provide opportunities for future research.

[6] UK SAP natural gas prices were downloaded from `www.nationalgrid.com/uk/Gas/Data`.

[7] Prices for Zeebrugge are from `www.apxendex.com` and were converted into kWh using 1 kWh=29.3072222 therm.

[8] The bond yield data were obtained from Reuters.

## 7. Discussion

We presented two novel techniques for estimation of a linear cointegration relationship for time-series. First, the method in section 3 allows estimation of a linear relationship between variables and we showed that our resulting algorithm is less likely to deliver a spurious relationship than the classical OLS–unit root testing approach. The benefit of our approach is that it results in a fast $O(T)$ algorithm for testing cointegration. A natural extension would be to consider a more complete Bayesian analysis and marginalise over all parameters in both the cointegration and random walk models.

Second, we developed a switching model to detect a relationship and regimes in which that relationship is a random walk. Whilst the model has an obvious limitation (shared by other existing intermittent cointegration models) it has the benefit of fast exact inference, scaling $O(T^2)$. A further natural extension would be to consider regimes in which the linear relationship can also vary. A full derivation of inference for the switching model is given in the supplementary material.

### Acknowledgements

We thank Ricardo Silva for useful discussions.

## References

Barber, D. *Bayesian Reasoning and Machine Learning.* Cambridge University Press, 2012.

Bracegirdle, C. and Barber, D. Switch-Reset Models : Exact and Approximate Inference. In *AISTATS*, volume 15. JMLR, 2011.

Dickey, D. A., Jansen, D. W., and Thornton, D. L. A primer on cointegration with an application to money and income. *Federal Reserve Bank of St. Louis Review*, (Mar):58–78, 1991.

Granger, C. W. J. Developments in the study of cointegrated economic variables. *Oxford Bulletin of Economics and Statistics*, 48(3):213–228, 1986.

Granger, C. W. J. and Newbold, P. Spurious regressions in econometrics. *Journal of Econometrics*, 2(2):111–120, 1974.

Gregory, A. W. and Hansen, B. E. Residual-based tests for cointegration in models with regime shifts. *Journal of Econometrics*, 70(1):99–126, January 1996.

Harris, R. and Sollis, R. *Applied Time Series Modelling and Forecasting.* Wiley, 2003.

Hatemi-J, A. Tests for cointegration with two unknown regime shifts with an application to financial market integration. *Emp. Economics*, 35(3):497–505, 2008.

Kim, J. Y. Inference on Segmented Cointegration. *Econometric Theory*, 19(4):620–639, 2003.

Koop, G. M., Strachan, R. W., Van Dijk, H., and Villani, M. Bayesian approaches to cointegration. In *The Palgrave Handbook of Theoretical Econometrics*, pp. 871–898. Palgrave Macmillan, 2006.

## A. Sequential inference of $\phi$

Given a sequence of 'observations' $\epsilon_{1:T}$ we describe here a method to sequentially update the posterior distribution $p(\phi|\epsilon_{1:T})$. An extension of this inference routine is used in the intermittent cointegration model in section 5. We place an improper uniform prior $p(\epsilon_1) = \mathcal{U}(\epsilon_1|\mathbb{R})$, representing our belief that $\epsilon_1$ may take any real value, and use a recursive update filtering routine—see for example Barber (2012).

Initially, we begin with the prior $p(\phi) = \mathcal{U}(\phi|\,(-1,1))$, and then update the distribution by finding the posterior after observing each $\epsilon_t$. For $\epsilon_1$, there is no update to make since $p(\phi|\epsilon_1) \propto p(\phi)\,p(\epsilon_1)$. Thereafter, $p(\phi|\epsilon_{1:t}) \propto p(\epsilon_t|\epsilon_{t-1},\phi)\,p(\phi|\epsilon_{1:t-1})$. This can be calculated analytically since, if $p(\phi|\epsilon_{1:t-1}) \propto p(\phi)\,\mathcal{N}(\phi|f_{t-1},F_{t-1})$, then[9] $p(\phi|\epsilon_{1:t}) \propto p(\phi)\,\mathcal{N}(\phi|f_t,F_t)$ where

$$f_t = \frac{f_{t-1}\sigma^2 + \epsilon_t\epsilon_{t-1}F_{t-1}}{\sigma^2 + \epsilon_{t-1}^2 F_{t-1}}, \quad F_t = \frac{\sigma^2 F_{t-1}}{\sigma^2 + \epsilon_{t-1}^2 F_{t-1}} \quad (1)$$

which on setting the initial[10] $f_2$, $F_2$ from the emission

$$\mathcal{N}(\epsilon_2|\phi\epsilon_1,\sigma^2) = \frac{1}{|\epsilon_1|}\mathcal{N}\left(\phi\middle|\frac{\epsilon_2}{\epsilon_1},\frac{\sigma^2}{\epsilon_1^2}\right) \quad (2)$$

defines a recursion for the parameters $f_t$, $F_t$. After completing the updates, the required posterior $p(\phi|\epsilon_{1:T})$ is given by a Gaussian distribution with mean $f_T$ and covariance $F_T$ truncated to the interval $(-1,1)$.

[9] Details of the derivation are omitted; recursions correspond to Kalman updates—see for example Barber (2012).

[10] In the unlikely event that $\epsilon_1 = 0$, the Gaussian term arises from the first non-zero $\epsilon_{t-1}$.